\documentclass[11pt]{article}
\usepackage{moriond,epsfig}

\def\etal{{\it et al.}}

\input babarsym

\newcommand{\gevcccc}{\ensuremath{{\mathrm{\,Ge\kern -0.1em V^2\!/}c^4}}\xspace}

\def\Kmaybestar {\ensuremath{K^{(*)}\xspace}}

\def\kll {\B\to\Kmaybestar\ellell\xspace}
\def\kmaybeee {\B\to\Kmaybestar\epem\xspace}

\def\kmaybemm {\B\to\Kmaybestar\mumu\xspace}

\def\mkpi {\ensuremath{m_{\kaon\pi}}\xspace}
\def\mll {\ensuremath{m_{\ell\ell}}\xspace}

\def\modekavgll {\ensuremath{B\to K\ellell}\xspace}

\def\modekstll {\ensuremath{B\rightarrow K^{*}\ellell}\xspace}

\begin{document}
\vspace*{4cm}
\title{\babar\, MEASUREMENTS ON $\kll$ RATES AND RATE ASYMMETRIES}

\author{ L. SUN (on behalf of the \babar\, collaboration)}

\address{Department of Physics, University of Cincinnati\\
345 Clifton Ct, Cincinnati OH 45221, USA}

\maketitle\abstracts{
Based on 471 million \BB pairs collected with the \babar\, detector
at the \pep2\, \epem collider, we perform a series of measurements 
on rare decays $B\to \Kmaybestar\ellell$, where $\ellell$ is
either $\epem$ or $\mumu$. The measurements include
total branching fractions, and partial branching fractions in 
six bins of di-lepton mass-squared. 
We also measure isospin asymmetries in the same six bins. 
Furthermore,
we measure direct $\CP$ and lepton flavor asymmetries for 
    di-lepton mass below and above the 
    $\jpsi$ resonance. Our measurements show good agreement with both Standard
    Model predictions and measurements from other experiments.
}

\section{Introduction}
\label{sec:intro}
\begin{figure}[b]
\begin{center}
\includegraphics[height=3cm]{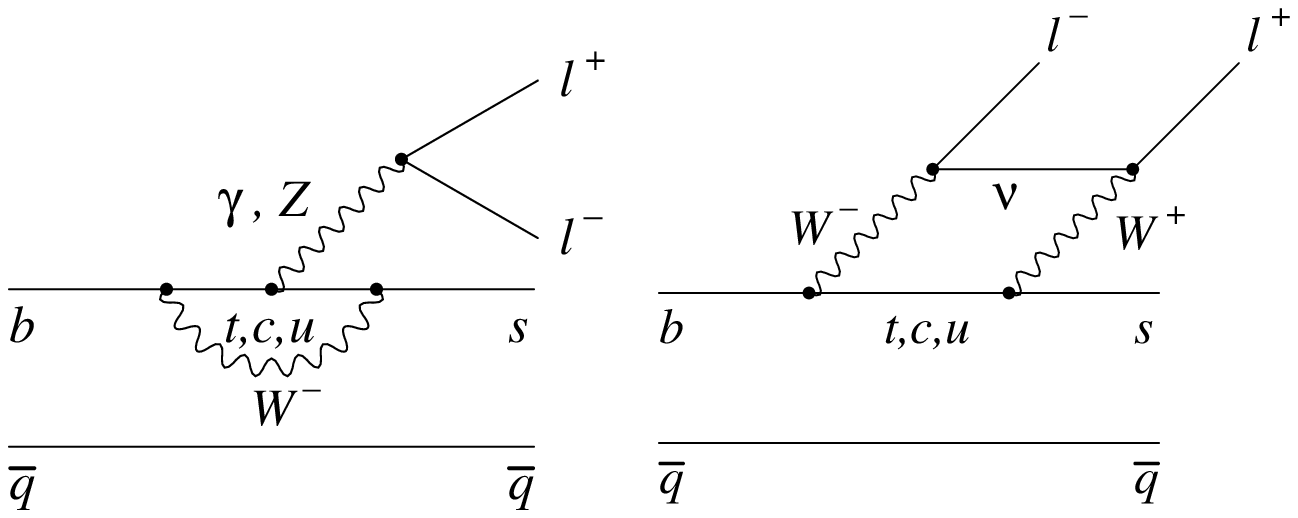}\\\vspace{5 mm}
\includegraphics[width=0.15\linewidth]{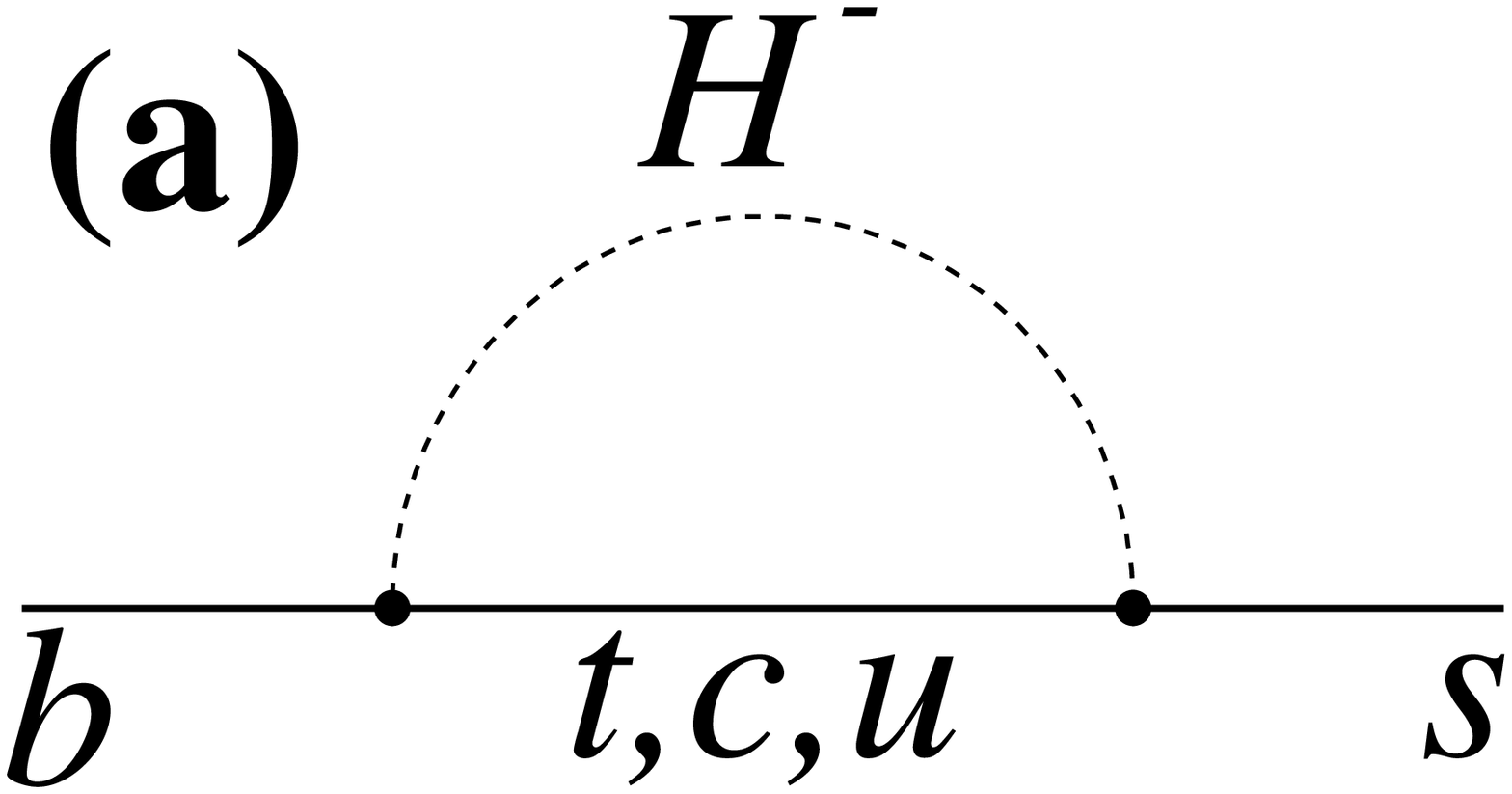}
\includegraphics[width=0.15\linewidth]{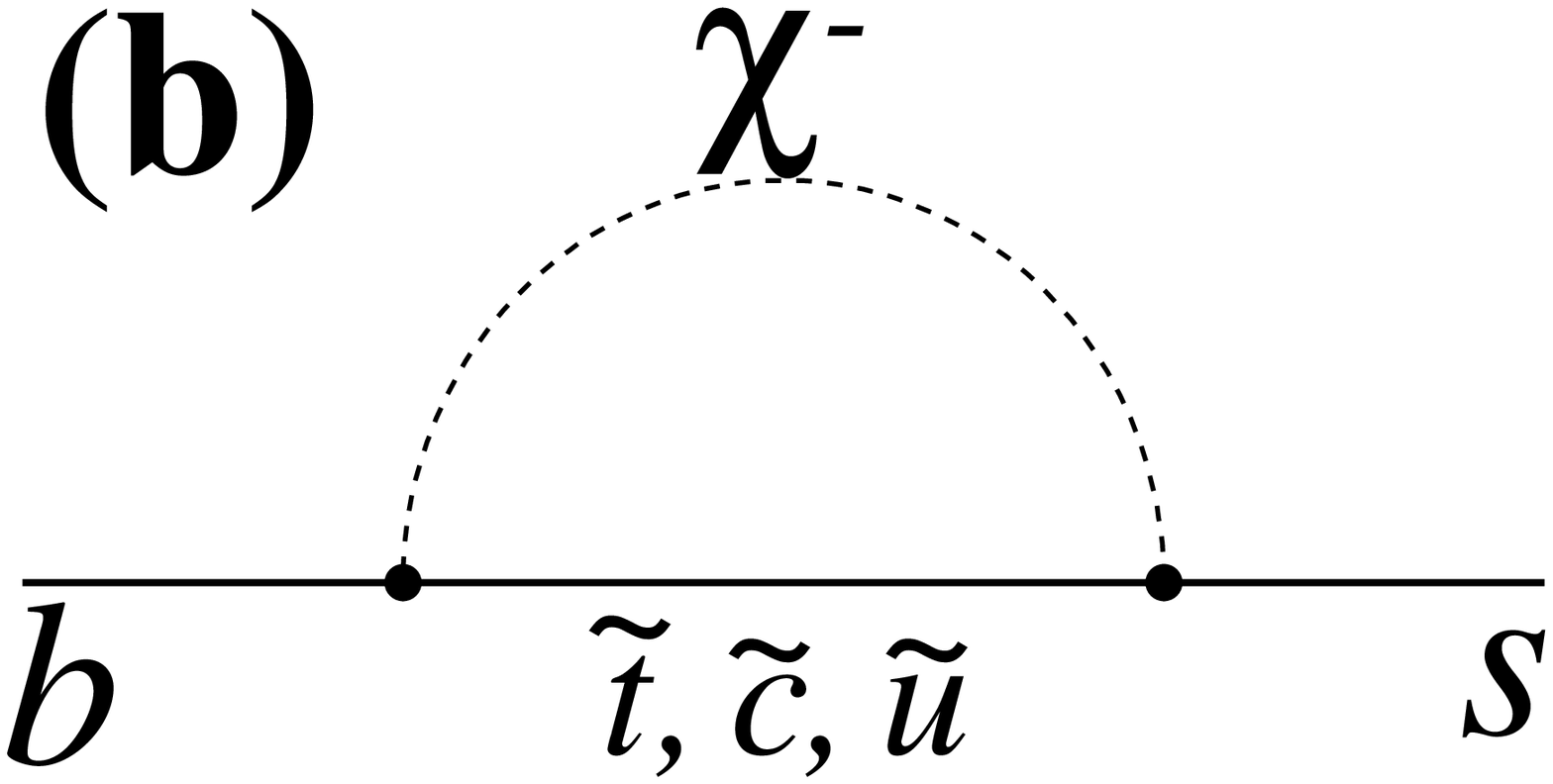}
\includegraphics[width=0.15\linewidth]{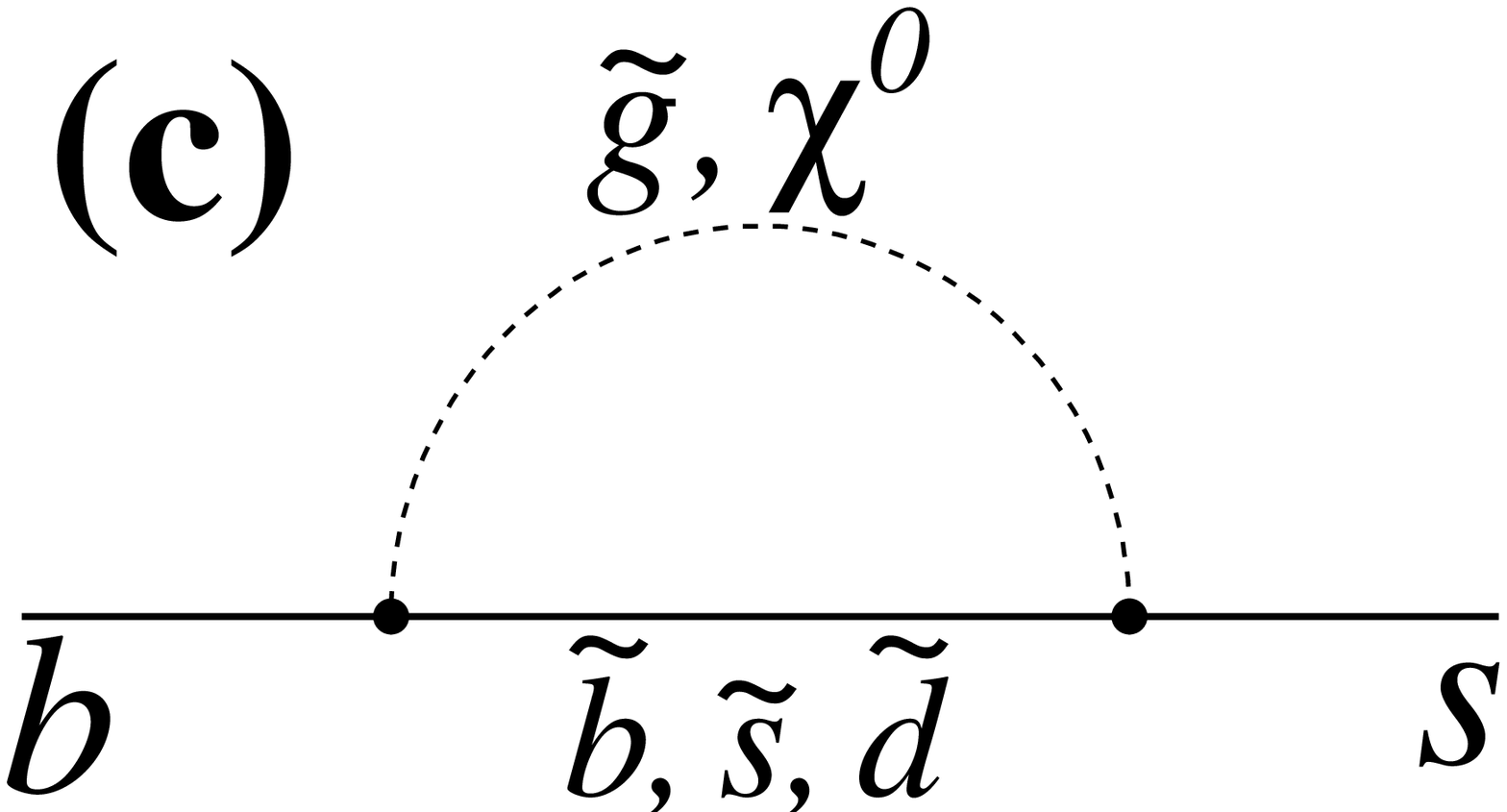}
\caption{Top: Lowest-order Feynman diagrams for $\btosll$ in the SM.
Bottom: Examples of new physics loop contributions to $\btosll$: (a) charged Higgs ($H^-$); (b) squark ($\tilde{t}, \tilde{c}, \tilde{u}$) and chargino ($\chi^-$); (c) squark ($\tilde{b}, \tilde{s}, \tilde{d}$) and gluino ($\tilde{g}$)/neutralino ($\chi^0$).}
\label{fig:sll_diagrams}
\end{center}
\end{figure}

The decays $\modekavgll$ and $\modekstll$~\cite{chargeconj} arise from
flavor-changing neutral-current (FCNC) $b\to s\ellell$ processes,
    which are forbidden at tree level in the Standard Model (SM).
These FCNC processes proceeds at lowest-order via $\gamma$/$Z$ penguin
and $W^+ W^-$ box diagrams~\cite{Buchalla} shown in Fig.~\ref{fig:sll_diagrams}.
New physics at the electro-weak scale may introduce new box and penguin diagrams
at the same order as the SM diagrams~\cite{Ali:2002jg}. Figure~\ref{fig:sll_diagrams} also show examples 
of these new physics loop processes. 

In the decays $\kll$, many observables are sensitive to new physics 
contributions. Due to poor knowledge of the $B\to\Kmaybestar$ form factors,
the theoretical predictions on decay rates possess large uncertainties. 
However most of the theoretical uncertainties cancel for the
ratios of these rates, the $B\to\Kmaybestar\ellell$ rate asymmetries 
can be particularly sensitive to new physics contributions 
due to better theoretical knowledge. By performing
the rate asymmetry measurements,
we are able to probe for new physics at the TeV scale~\cite{tevref}.

\section{Measurements}
\label{sec:meas}
The measurements are based on a data sample of 471 million $\BB$ pairs
collected at the \Y4S\, resonance
with the \babar\, detector~\cite{BaBarDetector} at the \pep2\, asymmetric-energy $\epem$ 
collider at the SLAC National Accelerator Laboratory. 
We reconstruct \kll signal events in eight final states with
an $\epem$ or $\mumu$ pair, and a $\KS$, $\Kp$, $\Kstarp(\to\KS\pip)$,
   or $\Kstarz(\to\Kp\pim)$, where a $\KS$ candidate
   is reconstructed in the $\pipi$
  final state. We also require selected $\Kstar$ candidates to have an invariant
  mass of $0.72<\mkpi<1.10$~\gevcc. 
We perform measurements in six bins of di-lepton mass squared $s\equiv\mll^2$:
$0.1\le s<2.0$~\gevcccc, $2.0\le s<4.3$~\gevcccc, $4.3\le s<8.1$~\gevcccc,
$10.1\le s<12.9$~\gevcccc, $14.2\le s<16.0$~\gevcccc, and $ s\ge 16.0$~\gevcccc.
The experimental details on event selection and signal extraction
are presented in Ref.~\cite{babarpaper}.

\begin{figure}[b!]
\begin{center}
\includegraphics[height=4.0cm]{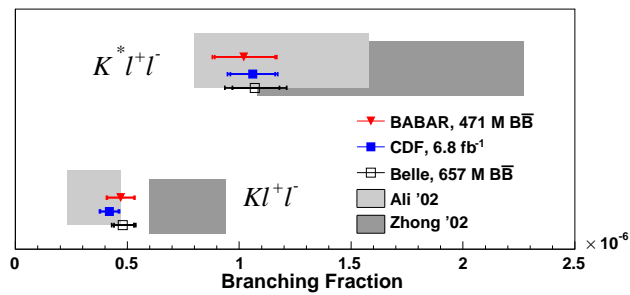}
\caption{Total branching fractions for the 
    $K \ell^+ \ell^-$ and  $K^* \ell^+ \ell^-$ modes compared
       with Belle~\protect\cite{belle09} and CDF~\protect\cite{cdf11} 
       measurements and with 
    predictions from the Ali~\etal~\protect\cite{Ali:2002jg}, 
and Zhong~\etal~\protect\cite{Zhong:2002nu} models. }
\label{fig:klltbf}
\end{center}
\end{figure}

\begin{figure}[b!]
\begin{center}
\includegraphics[height=6.0cm]{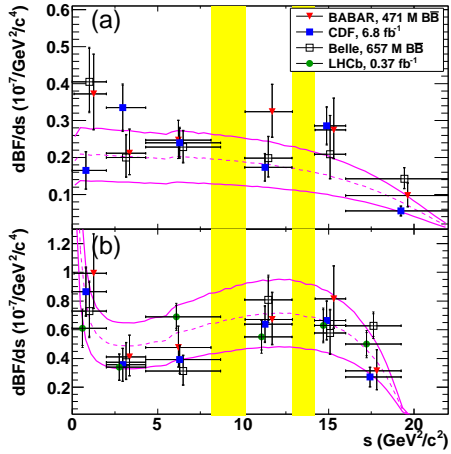}
\caption{Partial branching fractions for the (a) $K \ell^+ \ell^-$ and (b) $K^* \ell^+ \ell^-$ modes as a function of $s$ showing \babar\, measurements, Belle measurements~\protect\cite{belle09}, CDF measurements~\protect\cite{cdf11}, LHCb measurements~\protect\cite{lhcb11}, and the SM prediction from the Ali~\etal~model~\protect\cite{Ali:2002jg} with $B\to\Kmaybestar$ form factors~\protect\cite{ffmodels} (magenta dashed lines).
       The magenta solid lines show the theory uncertainties. The vertical 
       yellow shaded bands show the vetoed $s$ regions around the $J/\psi$ and $\psitwos$. }
\label{fig:kllpbf}
\end{center}
\end{figure}

We measure the total 
branching fractions for decays
\modekavgll and \modekstll
at $(4.7\pm0.6\pm 0.2) \times 10^{-7}$
and $(10.2_{-1.3}^{+1.4}\pm 0.5) \times 10^{-7}$, respectively.
Here, the first uncertainty is statistical, and the second is systematic.
Figure~\ref{fig:klltbf} show our total branching fraction results 
in good agreement with measurements from Belle~\cite{belle09} and CDF~\cite{cdf11}
and predictions from Ali~\etal~\cite{Ali:2002jg} and Zhong~\etal~\cite{Zhong:2002nu}.
Figure~\ref{fig:kllpbf} shows our results on $\kll$ partial branching fractions 
together with other recent experimental results from Belle~\cite{belle09}, CDF~\cite{cdf11}, and LHCb~\cite{lhcb11}. 
Our results
are also shown to be consistent with the predictions from Ali~\etal~\cite{Ali:2002jg}.

\begin{figure}[b!]
\begin{center}
\includegraphics[height=6cm]{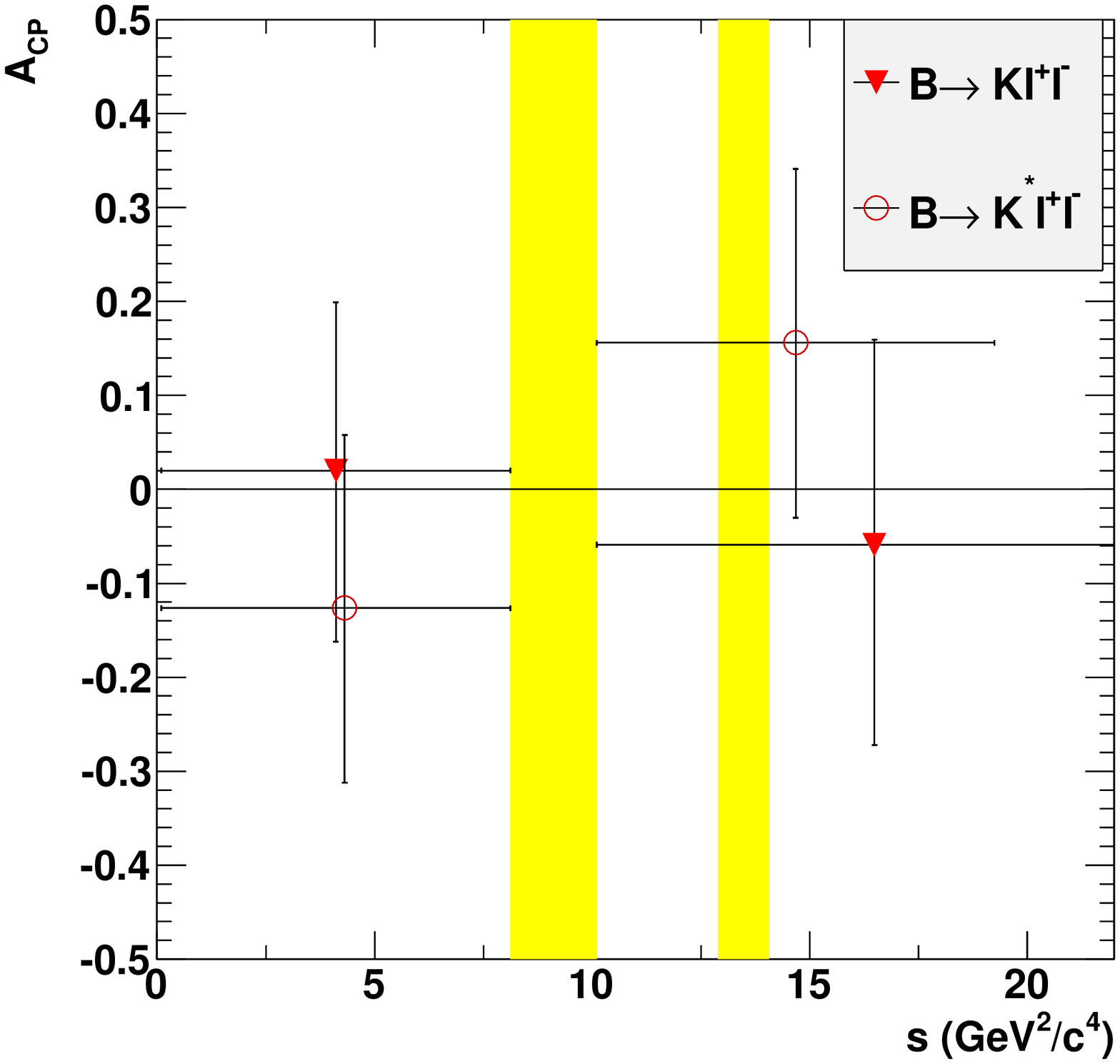}
\includegraphics[height=6cm]{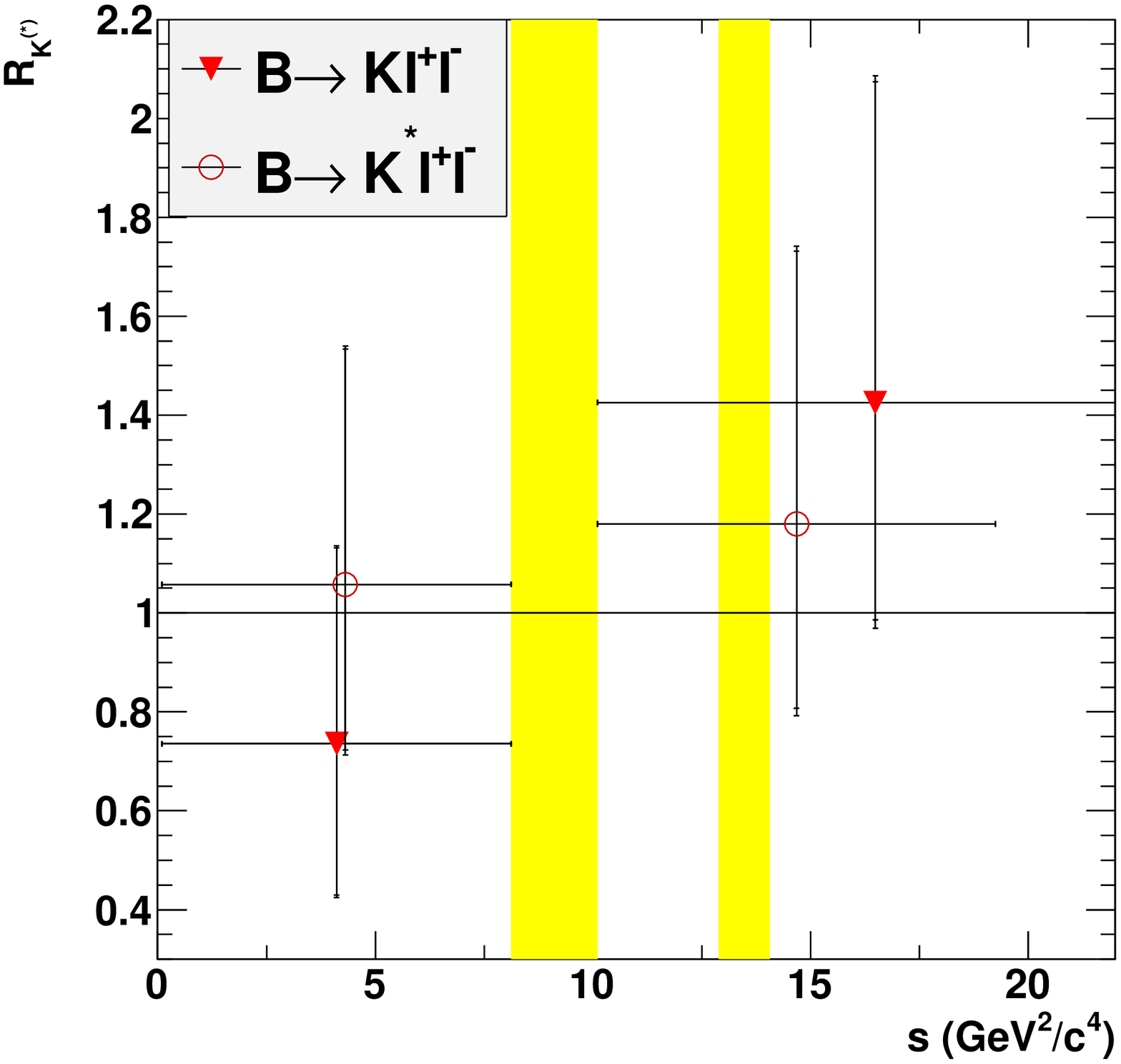}
\caption{(left) \CP asymmetries ${\cal A}_{CP}$ and (right) $R_{\Kmaybestar}$ for $K \ell^+ \ell^-$ modes and  $K^* \ell^+ \ell^-$ modes as a function of $s$. The vertical 
       yellow shaded bands show the vetoed $s$ regions around the $J/\psi$ and $\psitwos$.  }
\label{fig:acprk}
\end{center}
\end{figure}

The direct $\CP$ asymmetry
\begin{eqnarray}
{\cal A}_{\CP}^{\Kmaybestar} \equiv
\frac
{{\cal B}(\overline{B} \rightarrow \overline{K}^{(*)}\ellell) - {\cal B}(B \rightarrow K^{(*)}\ellell)}
{{\cal B}(\overline{B} \rightarrow \overline{K}^{(*)}\ellell) + {\cal B}(B \rightarrow K^{(*)}\ellell)}
\end{eqnarray}
is expected to be ${\cal O}(10^{-3})$ in the SM. However new physics at the electroweak weak scale
may bring in significant enhancement to ${\cal A}_{\CP}^{\Kmaybestar}$~\cite{CPnp}. 
The lepton flavor ratio
\begin{eqnarray}
{\cal R}_{\Kmaybestar} \equiv
\frac
{{\cal B}(\kmaybemm)}
{{\cal B}(\kmaybeee)}
\end{eqnarray}
is expected to be consistent with unity to within a few percent for $s>(2m_{\mu})^2$ in the SM~\cite{Hiller:2003js}. According to
two-Higgs-doublet models, the presence of a neutral Higgs boson at large $\tan\beta$ might 
increase ${\cal R}_{\Kmaybestar}$ by up to 10\%~\cite{Yan:2000dc}.
In Fig.~\ref{fig:acprk}, 
our ${\cal A}_{\CP}^{\Kmaybestar}$ and ${\cal R}_{\Kmaybestar}$ results 
below and above the $\jpsi$ resonance are shown to be in agreeement with the SM.

\begin{figure}
\begin{center}
\includegraphics[height=6cm]{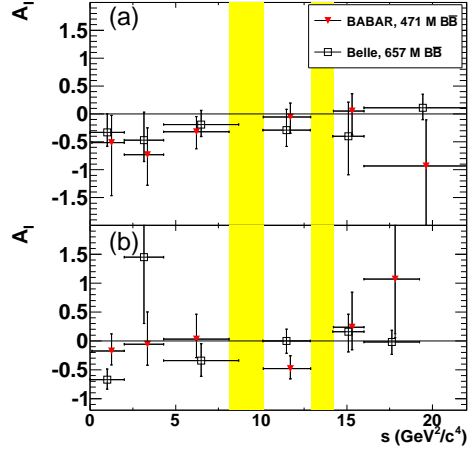}
\caption{Isospin asymmetry ${\cal A}_I$ for the (a) $K \ell^+ \ell^-$ and (b) $K^* \ell^+ \ell^-$ modes as a function of $s$, 
    in comparison to results from Belle~\protect\cite{belle09}. The vertical 
       yellow shaded bands show the vetoed $s$ regions around the $J/\psi$ and $\psitwos$. }
\label{fig:isospin}
\end{center}
\end{figure}

We also measure the $\CP$-averaged isospin asymmetry
\begin{equation}
{\cal A}^{K^{(*)}}_{I} \equiv
\frac
{{\cal B}(\Bz \to K^{(*)0}\ellell) - r_\tau {\cal B}(\Bp \to K^{(*)+}\ellell)} 
{{\cal B}(\Bz \to K^{(*)0}\ellell) + r_\tau {\cal B}(\Bp \to K^{(*)+}\ellell)},
\label{eq:aidef}
\end{equation}
\noindent
where $r_\tau \equiv \tau_{\Bz}/\tau_{\Bp}=1/(1.071\pm 0.009)$
is the ratio of $B^0$ and $B^+$ lifetimes~\cite{PDG}. 
In the SM, ${\cal A}^{K^{(*)}}_{I}$ is
expected to be small of a few percent. As $s\to 0$,
the SM expectation of ${\cal A}^{K^{*}}_{I}$ arrives at
its maximum of $+6\%$ to $+13\%$~\cite{isospin}. 
Figure~\ref{fig:isospin} shows our ${\cal A}^{K^{(*)}}_{I}$
results compared to the Belle results in the six $s$ bins.
In addition, in the low $s$ region ($0.10<s<8.12$~\gevcccc), we measure 
${\cal A}^{K}_I=  -0.58_{-0.37}^{+0.29}\pm0.02$ and 
${\cal A}^{\Kstar}_I=  -0.25_{-0.17}^{+0.20}\pm0.03$,
where the first uncertainty is statistical and the second is systematic.
Our ${\cal A}^{K}_{I}$ and ${\cal A}^{\Kstar}_{I}$ 
results are consistent with SM expectations of zero at 2.1$\sigma$ and 1.2$\sigma$, respectively.
These results also agree with the Belle measurements~\cite{belle09}.

\section{Summary}
\label{sec:summary}

In summary, we have performed measurements on total and partial 
branching fractions, direct $\CP$ asymmetries,
lepton-flavor ratios, and isospin asymmetries in the rare 
decays $\kll$ using the full \babar\, dataset of
471 million $B \bar B$ pairs. 
All our results are in good agreement with the SM predictions and those 
from Belle, CDF, and LHCb.
We notice negative isospin asymmetries in $\kll$ modes
at low $s$ values as seen by Belle.  

\section*{Acknowledgments}
The research is supported by the United States National Science Foundation
grants.
I also want to express my gratitude to the conference organizers for 
their great work.


\begin{thebibliography}{99}
\bibitem{chargeconj}
Charge conjugation is implied throughout except as explicitly
noted.
\end{thebibliography}
\end{document}